\documentclass{svproc}
\usepackage{graphicx}%
\usepackage{multirow}%
\usepackage{amsmath,amssymb,amsfonts}%
\usepackage{mathrsfs}%
\usepackage[title]{appendix}%
\usepackage{xcolor}%
\usepackage{textcomp}%
\usepackage{manyfoot}%
\usepackage{booktabs}%
\usepackage{algorithm}%
\usepackage{algorithmicx}%
\usepackage{algpseudocode}%
\usepackage{listings}%
\usepackage{url}

\def\orcidID#1{\unskip$^{[#1]}$}

\begin{document}
\mainmatter              
\title{Fast unbiased sampling of networks with given expected degrees and strengths}
\titlerunning{Fast unbiased sampling for the canonical configuration models}  
%
\author{Xuanchi Li\inst{1}\orcidID{0009-0007-0733-0829} \and Xin Wang\inst{1}\orcidID{0009-0006-7395-6988}
\and Sadamori Kojaku\inst{1}\orcidID{0000-0002-9414-6814}}
\authorrunning{Xuanchi Li et al.} 

\institute{
4400 Vestal Parkway E,
School of Systems Science and Industrial Engineering,
Binghamton University
P.O. Box 6000
Binghamton, NY 13902-6000\\
\email{skojaku@binghamton.edu}
}

\maketitle

\begin{abstract}
The configuration model is a cornerstone of statistical assessment of network structure.
While the Chung-Lu model is among the most widely used soft configuration models that preserve a prescribed degree sequence, it systematically oversamples edges between large-degree nodes, leading to inaccurate statistical conclusions.
Although the maximum entropy principle offers unbiased configuration models, its high computational cost has hindered widespread adoption,
making the Chung-Lu model an inaccurate yet persistently practical choice.
Here, we propose fast and efficient sampling algorithms for the maximum-entropy-based models by adapting the Miller-Hagberg algorithm. Evaluation on 103 empirical networks demonstrates a 10--100 times speedup, making theoretically rigorous configuration models practical and contributing to a more accurate understanding of network structure.
\end{abstract}

\section{Introduction}
\label{sec:introduction}

Networks are a mixture of random and non-trivial structures, which are often challenging to distinguish~\cite{colizza2006detecting,kojaku2018core,milo2002network,orsini2015quantifying}.
For example, apparently non-random structures such as rich-club~\cite{colizza2006detecting,zhou2004rich}, nestedness~\cite{lee2016network,payrató_borràs2019breaking}, and core-periphery structure~\cite{kojaku2018core,barucca2016disentangling} can emerge purely from degree heterogeneity. Configuration models, which generate random networks with prescribed degree sequences~\cite{chung2002connected,fosdick2018configuring}, provide a rigorous framework for statistical testing against null models, serving as cornerstones for network analysis~\cite{cimini2021reconstructing,mastrandrea2014enhanced,milo2002network,picciolo2022weighted,aldecoa2013surprise,kojaku2018generalised,lancichinetti2010statistical}.


The Chung-Lu model~\cite{chung2002connected} is the most widely used \emph{soft} configuration model that generates networks with a given degree sequence on expectation (we refer to it simply as the configuration model hereinafter).
A critical limitation of the Chung-Lu model is that it oversamples edges between large-degree nodes~\cite{squartini2015unbiased,mastrandrea2014enhanced,sayama2018combinatorial,maslov2004detection}.
While this bias is well-known and often considered negligible in sparse networks~\cite{chung2002connected,newman2006finding}, the bias is consequential in network analysis because large-degree nodes are a key determinant of network structure.

The systematic bias in the Chung-Lu model has motivated several alternative approaches. Combinatorial methods have been proposed that calculate edge probabilities through exact counting of network configurations, particularly for dense networks~\cite{sayama2018combinatorial}.
While combinatorial methods primarily focus on undirected, unweighted networks, the maximum entropy (MaxEnt) models for networks offer a flexible, theoretically rigorous framework that can be extended to different types of networks, including weighted, directed, and bipartite networks~\cite{park2004statistical,squartini2015unbiased,vallarano2021fast}. Yet, MaxEnt models remain far from widespread adoption in practice due to two critical computational bottlenecks: parameter inference and network sampling. The inference bottleneck requires solving a non-linear optimization problem with at least $2N$ variables, resulting in computational complexity exceeding ${\cal O}(N^2)$. This challenge has recently been addressed through efficient solvers~\cite{parisi2020faster,vallarano2021fast}.
However, the sampling bottleneck remains unresolved, with the brute force algorithm being the commonly used option to sample networks from MaxEnt models. The brute force approach evaluates edge probabilities for all possible node pairs, requiring ${\cal O}(N^2)$ probability computations, resulting in computational complexity that becomes prohibitive for large-scale network analysis. This persistent gap between statistical rigor and computational feasibility makes the biased Chung-Lu model an inaccurate but the only viable option, hence perpetuating systematic errors in network analysis.

We introduce a fast sampling algorithm for configuration models based on the maximum entropy principle. Building on the rejection sampling proposed by Miller and Hagberg~\cite{miller2011efficient}, we adapt their method to work with the MaxEnt models by generating a set of candidate edges from an efficient proposal distribution, then selectively accepting these candidates to match the target probabilities. We develop two variants of the algorithm: one for unweighted networks and one for weighted networks. While the unweighted variant has been implemented as an experimental feature in igraph~\cite{igraph_games_section4.4}, our work provides the first systematic evaluation of its performance and extends the approach to weighted networks. The resulting algorithms maintain statistical correctness while achieving dramatic computational speedup, making theoretically rigorous configuration models practical for large network analysis and contributing to a more accurate understanding of the structure induced by degree heterogeneity.
A Python implementation of the proposed sampling algorithms can be found at \url{https://github.com/EKUL-Skywalker/fastmaxent}.

\section{Methods}
\label{sec:methods}
\subsection{Non-negligible bias of the Chung-Lu model}

The Chung-Lu model systematically oversamples edges between large-degree nodes. This systematic bias is well-known, and the impact of the bias in network analysis has been regarded as negligible~\cite{chung2002connected,newman2006finding}. We argue that the bias in the Chung-Lu model can overturn the results of network analysis by presenting a simple example network.

\begin{figure}
    \centering
    \includegraphics[width=0.8\textwidth]{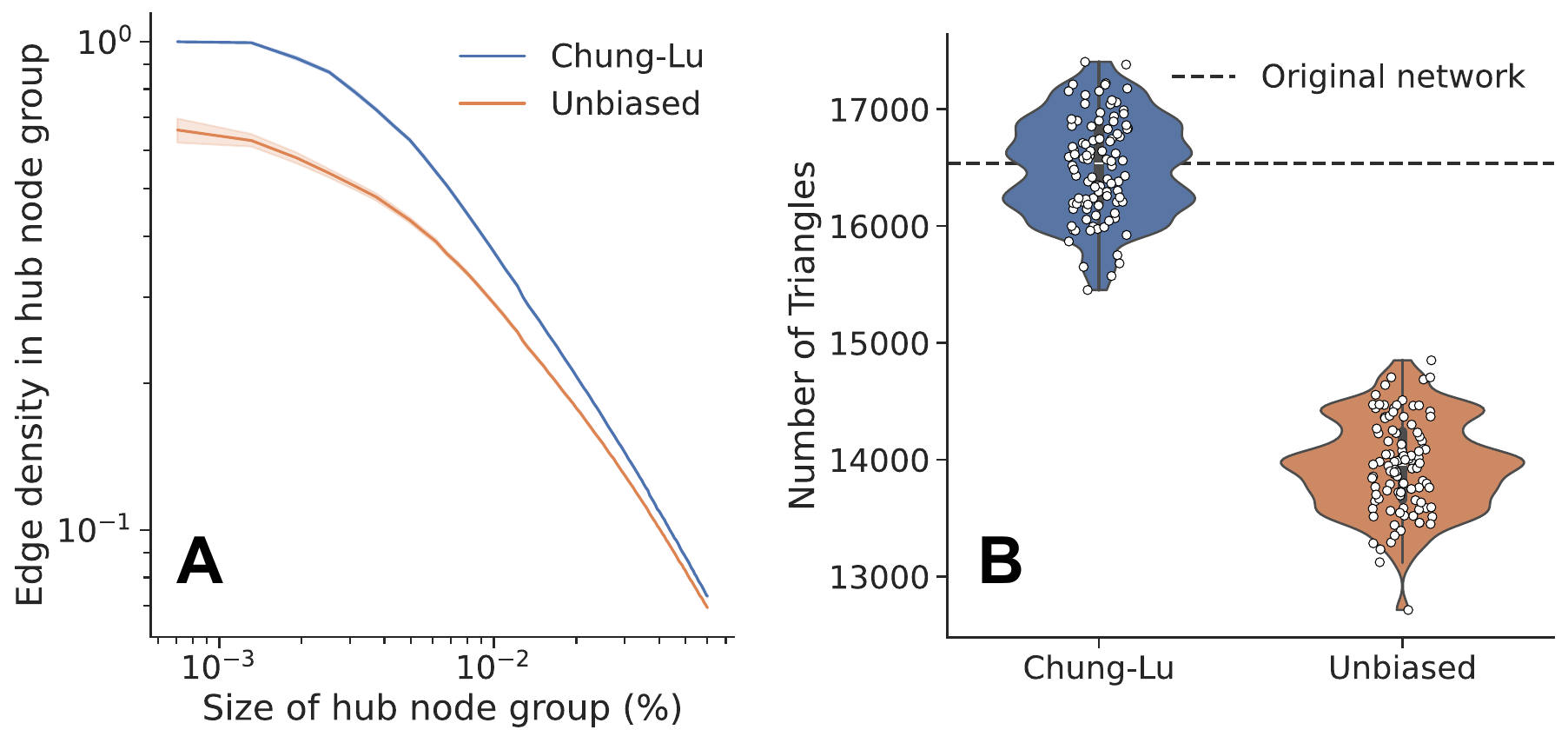}
    \caption{Structural bias in the Chung-Lu model. We use the Holme-Kim network of 5,000 nodes.
    {\bf A}: The density of edges within the largest-degree nodes as a function of group size $\alpha$. {\bf B}: The distribution of triangle counts in the sampled networks across 100 realizations.}
    \label{fig:chung-lu-bias}
\end{figure}

We generate a network of 5,000 nodes with both power-law degree distributions and high clustering coefficients using the Holme-Kim model~\cite{holme2002growing}, with 10 edges per node and a clustering probability of 0.1.
We then measure the density of edges within the group of the top $\alpha$\% largest-degree nodes (Fig.~\ref{fig:chung-lu-bias}A).
For a small $\alpha$, the density of edges among the highest-degree nodes is markedly greater in the Chung-Lu model than in the unbiased configuration model based on maximum entropy~\cite{garlaschelli2008maximum,mastrandrea2014enhanced}. In other words, the Chung-Lu model generates a rich-club structure~\cite{colizza2006detecting,zhou2004rich}, where large-degree nodes are more densely connected than expected by random chance, even though it should generate networks without any structure except the degree sequence.

This oversampling bias stems from a fundamental misapplication of the Chung-Lu model.
The Chung-Lu model generates edges between nodes $i$ and $j$ according to a Poisson distribution with rate parameter $\lambda_{ij}$ given by~\cite{chung2002connected} $\lambda_{ij} = k_i k_j/2M$,
where $k_i$ is the expected degree of node $i$, and $M = \sum_{i=1}^N k_i/2$ is the total expected number of edges.
Critically, $\lambda_{ij}$ represents the \emph{expected number of edges} between nodes $i$ and $j$, and hence should not be confused with the probability.
The original model, therefore, samples \emph{multigraphs} where multiple edges can appear between the same pair of nodes, which are often expressed as the weight of the edge.
However, the model is commonly misapplied as a null model for \emph{unweighted} networks by treating $\lambda_{ij}$ as an edge probability.
This requires the assumption that $\lambda_{ij} \ll 1$ for all node pairs, which fails systematically in degree-heterogeneous networks.

The ramifications of this bias extend beyond edge density and potentially mislead statistical characterizations of networks (Fig.~\ref{fig:chung-lu-bias}B).
When counting triangles in the Holme-Kim network, which by design contains more triangles than random, the Chung-Lu model generates a comparable number of triangles to the Holme-Kim network, incorrectly accepting the null hypothesis that triangles are not significant motifs.
In contrast, the unbiased configuration model samples fewer triangles and correctly rejects the null hypothesis, leading to opposite conclusions about network motifs.

\subsection{Undirected Binary Configuration Model (UBCM)}

The systematic bias of the Chung-Lu model has been recognized for a long time and has been theoretically addressed based on the principle of maximum entropy~\cite{garlaschelli2008maximum,mastrandrea2014enhanced}.
The principle of maximum entropy provides a framework for specifying an unbiased configuration model by maximizing the Shannon entropy subject to specific topological constraints~\cite{park2004statistical}. For the configuration model, the maximum entropy principle yields the \emph{undirected binary configuration model} (UBCM), with the edge probability being a sigmoid function given by~\cite{garlaschelli2008maximum,squartini2015unbiased,vallarano2021fast}:
\begin{align}
    p^{\text{UBCM}}_{ij} = \frac{\exp(-\alpha_i -\alpha_j)}{1 + \exp(-\alpha_i -\alpha_j)},
\end{align}
where $\alpha_i$ and $\alpha_j$ are node-specific parameters determined by maximum likelihood estimation to preserve the expected degree sequence~\cite{vallarano2021fast,mastrandrea2014enhanced}.

While the UBCM provides a theoretically unbiased framework for unweighted network generation, its practical implementation faces two critical computational challenges that have limited its widespread adoption. The first challenge is \emph{parameter inference}, which requires solving a non-linear optimization problem to estimate $\alpha_i$ through maximum likelihood estimation (MLE). This involves solving a system of equations with computational complexity ${\cal O}(N^2)$, though recent work introduces efficient solvers and models to reduce inference time~\cite{gabrielli2019grand,parisi2020faster,vallarano2021fast}.

The second and more persistent challenge is \emph{network sampling}. The traditional brute-force sampling approach evaluates the edge probability $p^{\text{UBCM}}_{ij}$ for every pair of nodes, which is impractical for sampling large networks. For example, for a network with $N = 10,000$ nodes, approximately 50 million probability evaluations are required. Furthermore, random networks are often used as null models for network analysis, requiring the generation of hundreds or thousands of network instances in an ensemble for robust statistical inference, multiplying this computational burden accordingly. Therefore, this sampling bottleneck remains the primary obstacle preventing the adoption of theoretically rigorous configuration models in practice.

\subsection{Miller-Hagberg Algorithm for the Chung-Lu model}

\begin{figure}[tb!]
    \centering
    \includegraphics[width=\textwidth]{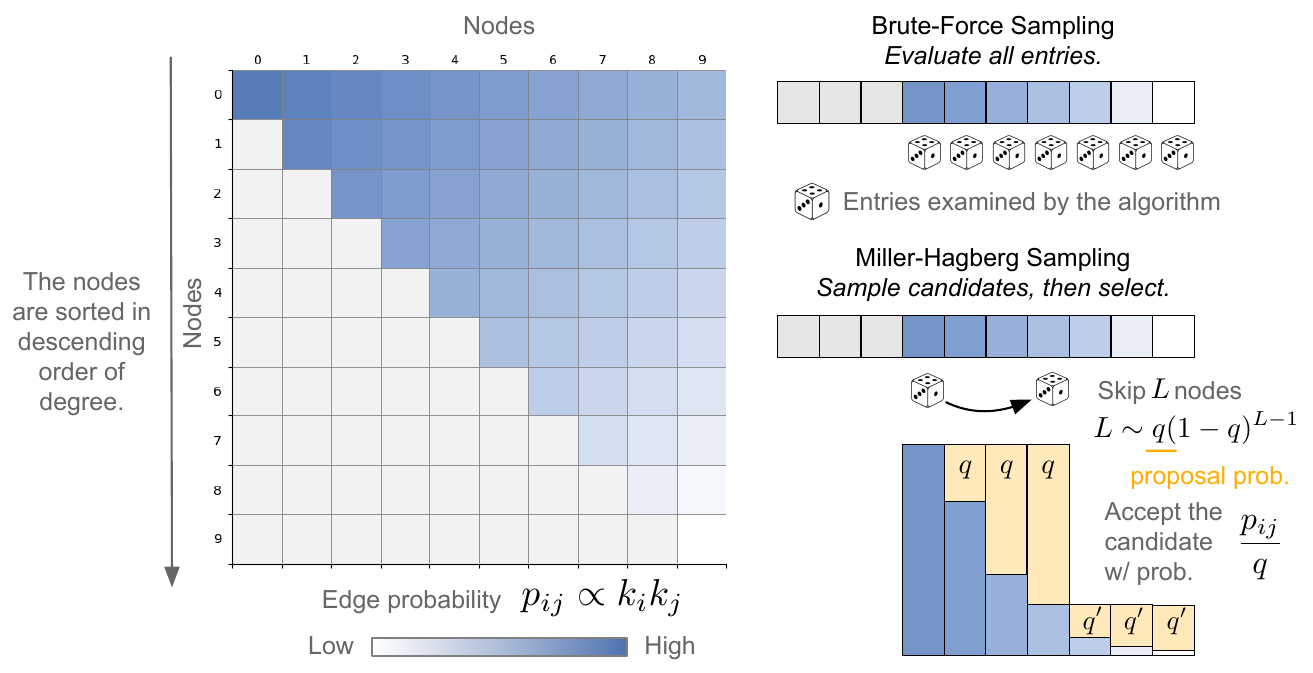}
    \caption{
        Schematic illustration of the Miller-Hagberg (MH) algorithm for the Chung-Lu model.
        In the Chung-Lu model, the edge probabilities $p_{ij}$ are monotonically decreasing with respect to the node degrees.
        The MH algorithm exploits this structure by proposing a neighbor with probability $q \geq p_{ij}$ and accepting candidates with probability $p_{ij}/q$.
        Proposing the next candidate neighbor after skipping over $L$ nodes follows a geometric distribution $q(1-q)^{L-1}$.
        By sampling $L$ from the geometric distribution, the algorithm \emph{avoids having} to evaluate all nodes.}
    \label{fig:schematic-mh}
\end{figure}

The Miller-Hagberg (MH) algorithm is an efficient sampling algorithm for the Chung-Lu model, and its core steps serve as the foundation for our proposed sampling algorithms for MaxEnt models. To understand the MH algorithm intuitively, let us start with a simple case where all nodes have identical degree $k$. In this scenario, the Chung-Lu model is equivalent to the Erdős-Rényi model. Consider a focal node $i$ that needs to find its neighbors among the remaining $N-1$ nodes.
While brute-force sampling goes through every node in order and evaluates the edge probabilities, the MH algorithm avoids this exhaustive evaluation by asking
``\emph{how many nodes can we skip before finding the next neighbor?}'' In other words, the MH algorithm evaluates the probability $p^{\text{Geo}}(L;p)$ that $L$ subsequent nodes do \emph{not} have edges with node $i$. The probability $p^{\text{Geo}}(L;p)$ follows a geometric distribution (i.e., $p^{\text{Geo}}(L;p) = (1-p)^{L}p$) because each node has the same edge probability $p$.
By noting that $pN^2 \simeq M$, the time needed for generating an Erdős-Rényi network is ${\cal O}(M)$ instead of ${\cal O}(N^2)$.

The idea of skipping nodes can be extended to the case where nodes have heterogeneous degrees (Fig.~\ref{fig:schematic-mh}). Consider node $i$ with degree $k_i$ seeking connections to other nodes sorted by decreasing degree: $k_1 \geq k_2 \geq \cdots \geq k_N$. In the Chung-Lu model, the edge probabilities are also sorted in decreasing order: $p_{i1} \geq p_{i2} \geq \cdots \geq p_{iN}$. The MH algorithm exploits this monotonic decrease through a \emph{sequential rejection sampling technique}: starting from the highest-degree candidate $j=1$, we set the proposal probability $q = p_{i1}$ and use the geometric distribution $p^{\text{Geo}}(L;q)$ to skip over nodes to find a potential neighbor $j$.
However, the proposal probability $q$ is larger than the true edge probability $p_{ij}$.
Rejection sampling corrects this discrepancy by accepting the candidate with probability $p_{ij}/q$.
In other words, node $j$ is sampled as a neighbor of node $i$ when it is sampled by the proposal probability $q$ and accepted by rejection sampling with probability $p_{ij}/q$, resulting in the correct probability $q \times p_{ij}/q = p_{ij}$.
Once node $j$ is evaluated, the proposal probability $q$ is updated to $q=p_{ij}$, and the next candidate $j+L$ is sampled, where $L$ is sampled from $p^{\text{Geo}}(L;q)$. This process is repeated until all nodes are evaluated.
This sequential rejection sampling technique with adaptive proposal probabilities reduces the expected number of evaluations to be nearly proportional to the number $M$ of edges.

\subsection{Sampling Algorithm for the UBCM}
\label{sec:proposed_sampling_algorithm_for_ubcm}

It is straightforward to adapt the Miller-Hagberg algorithm to the UBCM, which has been done in the igraph package~\cite{igraph_games_section4.4}.
While the exact probability formulas differ between the Chung-Lu model and the UBCM, both models share a crucial feature that enables efficient sampling: nodes can be ordered to create monotonically decreasing edge probability sequences. In the Chung-Lu model, we can create a decreasing probability sequence $p_{i1} \geq p_{i2} \geq \cdots \geq p_{iN}$ by sorting nodes in descending order of node degree $k_i$.
For the UBCM, a decreasing probability sequence can be created by sorting the nodes in \emph{ascending order} of $\alpha_i$ because lower $\alpha_i$ values yield higher edge probabilities.
The step-by-step description of the algorithm is as follows:
(i) We sort all nodes $i$ in ascending order of $\alpha_i$ (i.e., $\alpha_1 \leq \alpha_2 \leq \cdots \leq \alpha_N$).
(ii) We then go through each node $i$ in this order with the initial proposal probability $q = p^{\text{UBCM}}_{i,i}$.
(iii) For each focal node $i$, we sample a candidate neighbor $j$ ($j>i$ for undirected networks) by skipping $L$ nodes, where $L$ is sampled from the geometric distribution $p^\text{Geo}(L;q)$, and accept the candidate with probability $p^{\text{UBCM}}_{i,j}/q$. The proposal probability $q$ is then updated to $q=p^{\text{UBCM}}_{i,j}$, and this step is repeated until no candidate is left.
(iv) Continue the process with the next candidate. If we reach the end of the candidate list (e.g., $j+L > N$), move to the next node $i+1$.
While the sampling algorithm is implemented in the igraph package, it is still in the experimental stage, and the accuracy and speed are not yet fully tested. Thus, we implemented it ourselves and used it for our simulations. The code is included in our Python package.

\subsection{Sampling algorithm for weighted configuration model}
\label{sec:proposed_sampling_algorithm_for_weighted_configuration_model}

\begin{algorithm}[bt]
\caption{Undirected Enhanced Configuration Model (UECM)}
\label{alg:UECM}
\begin{algorithmic}[1]
\State \textbf{Input:} $(\alpha_1, \ldots, \alpha_N)$ and $(\beta_1, \ldots, \beta_N)$
\State \textbf{Output:} Edge list $\mathcal{E}$
\State Sort nodes by increasing $\alpha_i + \beta_i$: $\alpha_{\pi(1)} + \beta_{\pi(1)} \leq \cdots \leq \alpha_{\pi(N)} + \beta_{\pi(N)}$
\State Initialize empty edge list $\mathcal{E} \gets \{\}$
\For{$i = 1$ to $N$}
    \State $\beta_{\min} \gets$ minimum $\beta$ value among remaining unprocessed nodes
    \State $j \gets i + 1$, $q \gets p^{\text{UECM}}_{\pi(i)\pi(i)}(\beta_{\min})$
    \While{$j \leq N$}
        \State $p \gets p^{\text{UECM}}_{\pi(i)\pi(j)}$
        \If{$\text{Uniform}(0,1) < p/q$}
            \State $\mathcal{E} \gets \mathcal{E} \cup \{(\pi(i), \pi(j), w)\},\;\text{where}\; w\sim \text{Geometric}(1 - e^{-\left(\beta_{\pi(i)} + \beta_{\pi(j)}\right)})$
        \EndIf
        \State $q \gets p^{\text{UECM}}_{\pi(i)\pi(j)}(\beta_{\min})$
        \State $j \gets j + \text{Geometric}(q)$
    \EndWhile
\EndFor
\State \textbf{Return} $\mathcal{E}$
\end{algorithmic}
\end{algorithm}

For weighted networks, the principle of maximum entropy enables the Undirected Enhanced Configuration Model (UECM) to preserve the degree sequence along with the sequence of node strengths (i.e., the sum of the weights of the edges pertaining to each node).
Specifically, the probability that two nodes $(i,j)$ have an edge with weight $w_{ij}$ is given by~\cite{mastrandrea2014enhanced,vallarano2021fast,parisi2020faster}
\begin{align}
    \label{eq:P_A_UECM}
    p^{\text{UECM}}(w_{ij} = w) = p^{\text{UECM}}_{ij} \cdot p^{\text{UECM}}(w_{ij} = w \vert w_{ij} > 0),
\end{align}
where $p^{\text{UECM}}_{ij}$ governs the existence of an edge and $p^{\text{UECM}}(w_{ij} = w \vert w_{ij} > 0)$ determines the weight given an edge~\cite{mastrandrea2014enhanced,vallarano2021fast,parisi2020faster}, i.e.,
\begin{align}
    p^{\text{UECM}}_{ij} &= \frac{\exp(-\alpha_i -\alpha_j - \beta_i -\beta_j)}{1 - \exp(-\beta_i -\beta_j) + \exp(-\alpha_i -\alpha_j - \beta_i -\beta_j)}, \label{eq:P_A_UECM_w_gt_0} \\
    p^{\text{UECM}}(w_{ij} = w \vert w_{ij} > 0) &= \exp(-\beta_i - \beta_j)^{w-1} [1 - \exp(-\beta_i -\beta_j)]. \label{eq:P_A_UECM_w_gt_0_cond}
\end{align}
Here, $\alpha_i$ and $\beta_i$ are node-specific parameters: $\alpha_i$ controls the propensity to form connections (degree constraint), while $\beta_i$ controls the strength of those connections (strength constraint). The conditional weight distribution follows a geometric distribution with parameter $\exp(-\beta_i - \beta_j)$.

Adapting the MH algorithm for the UECM is challenging because the edge probabilities now depend on both the $\alpha_i$ and $\beta_i$ parameters, which do not create a single node ordering that generates decreasing edge probabilities.
We circumvent this problem using the following key idea: we work with an upper bound $\hat{p}^{\text{UECM}}_{ij}$ of the true probability that can create decreasing edge probabilities, and then correct for the error using rejection sampling.
Let us revisit Eq.~\eqref{eq:P_A_UECM_w_gt_0} with numerator $\exp(-\alpha_i -\alpha_j - \beta_i -\beta_j)$ and denominator $1 - \exp(-\beta_i -\beta_j) + \exp(-\alpha_i -\alpha_j - \beta_i -\beta_j)$.
Notice that the numerator and denominator share the same sum, $-\alpha_i-\alpha_j-\beta_i-\beta_j$, in common.
What prevents us from creating decreasing edge probabilities is the second term of the denominator, $\exp(-\beta_i -\beta_j)$, and we replace it with $\exp(-\beta_{\min})$, where $\beta_{\min} = \min_{\ell, \ell'} \beta_\ell + \beta_{\ell'}$ is the minimum sum of $\beta$ values among unprocessed nodes $\ell$ and $\ell'$.
Since $\exp(-\beta_{\min}) \geq \exp(-\beta_i -\beta_j)$, this leads to the upper bound probability
\begin{align}
    \hat{p}^{\text{UECM}} _{ij} = \frac{\exp(-\alpha_i -\alpha_j - \beta_i -\beta_j)}{1 - \exp(-\beta_{\min}) + \exp(-\alpha_i -\alpha_j - \beta_i -\beta_j)}.
\end{align}
The minimum sum $\beta_{\min}$ can be computed efficiently by pre-sorting $\beta_i$ values in ascending order and taking the first two $\beta$ values of the unprocessed nodes.
The step-by-step procedure is described in Algorithm~\ref{alg:UECM}.

\subsection{Extension to other network types}
\label{sec:extension_bipartite_directed}

Our sampling algorithm naturally extends to bipartite networks.
We denote parameters for one node set as $\alpha^+, \beta^+$ and for the other node set as $\alpha^-, \beta^-$.
We then go through each node in the $+$ node set and then go through each node in the $-$ node set to sample edges.
The order of the nodes for the $+$ node set and the $-$ node set is sorted according to $\alpha^+_i,\beta^+_i$ and $\alpha^-_j,\beta^-_j$, respectively.

The sampling algorithm for bipartite networks can be used to generate directed networks as follows. Given a directed network, we create two nodes per node: one corresponding to the \emph{in} nodes and the other corresponding to the \emph{out} nodes, with one node type placed in one part of the bipartite network and the other node type in the other part. An edge from node $i$ to node $j$ in the directed network corresponds to an edge between out-node $i$ and in-node $j$ in the bipartite representation. This transformation allows us to apply the bipartite sampling algorithm to generate directed networks while preserving both in-degree and out-degree sequences. In the same vein, our algorithm can be applied to hypergraphs by representing them as incidence graphs.

\section{Results}
\label{sec:results}

\begin{figure}[tb!]
    \centering
    \includegraphics[width=\textwidth]{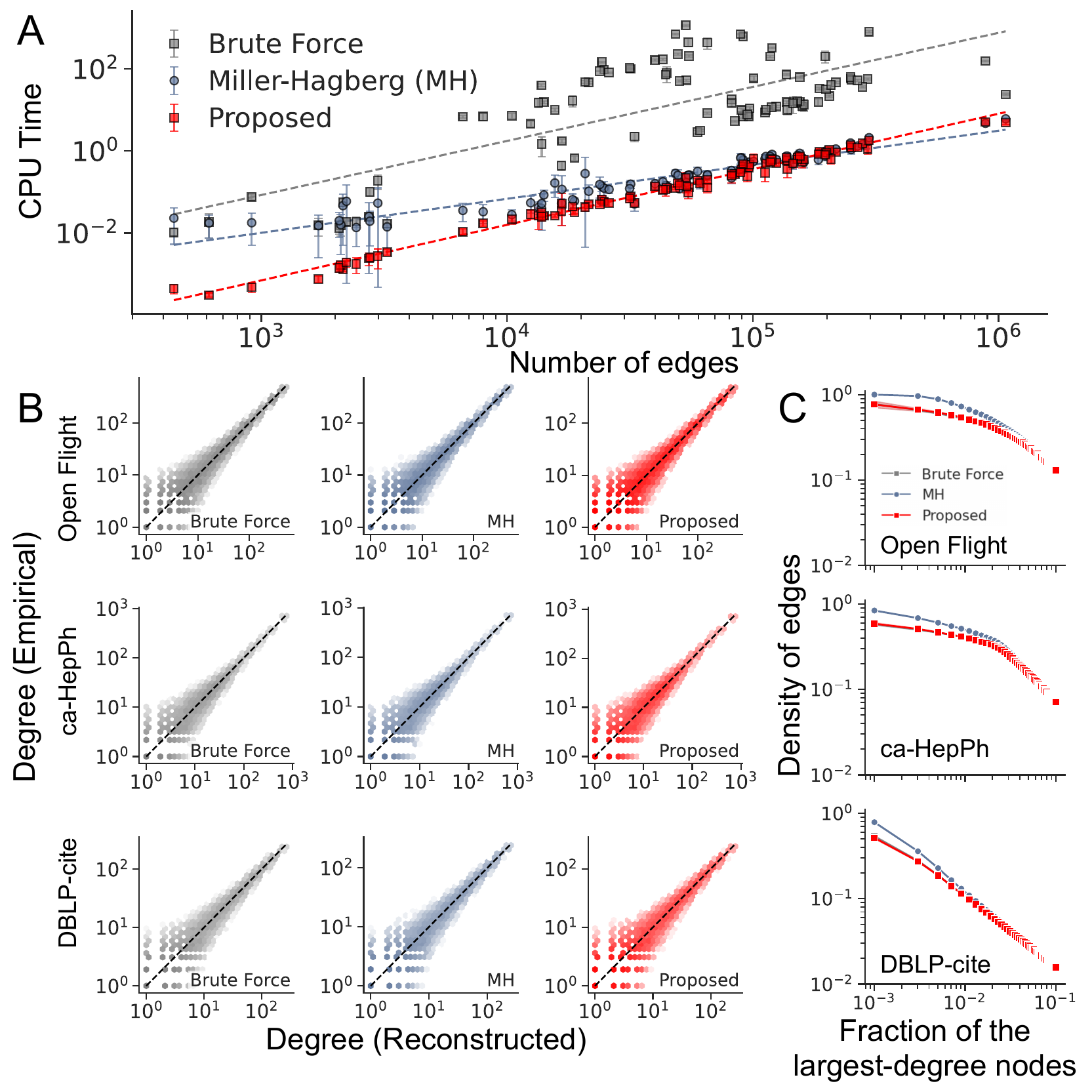}
    \caption{
    Results for the unweighted networks.
    {\bf A}: CPU Time as a function of the number $M$ of edges.
    {\bf B}: The joint probability distribution of the reconstructed and empirical degree sequences.
    We show the results for three empirical networks as representative (Open Flight, ca-HepPh, and DBLP-cite).
    {\bf C}: The density of edges within the group of the largest-degree nodes.
    }
    \label{fig:result_unweighted}
\end{figure}

We analyzed 93 unweighted networks across domains and sizes, removing self-loops and multi-edges (for network statistics, see Ref.~\cite{aiyappa2025implicit}). We also used 10 weighted networks, with edge weights rounded to integers as required by the Chung-Lu and UECM models~\cite{vallarano2021fast}.
To cover a wider range of network sizes, we converted the 93 unweighted networks to weighted networks by assigning one plus the number of common neighbors as edge weights, with one added to avoid zero weights.
Parameters were inferred using \texttt{NEMtropy}~\cite{vallarano2021fast} with quasi-Newton optimization. Parameter inference time was excluded from sampling CPU measurements. We compared our algorithm against Brute Force (BF) sampling for MaxEnt models, MH for Chung-Lu (unweighted only), and strength-based sampling for Chung-Lu (weighted only). BF evaluates all node pairs using the UBCM~\cite{vallarano2021fast}. MH samples networks using Chung-Lu probabilities~\cite{miller2011efficient}.
Strength-based sampling first samples the total number of edges from a Poisson distribution with a mean equal to the sum of edge weights. Then, we attach each edge to a node pair with probability proportional to their node strengths.
Each algorithm ran 40 times per network, measuring CPU time excluding wait time.
We refer to our algorithm as FastMaxEnt.

\subsection{Unweighted networks}
\label{sec:result_unweighted}

The FastMaxEnt algorithm samples networks as efficiently as the MH method while being as accurate as the BF method (Fig.~\ref{fig:result_unweighted}).
For CPU time, the FastMaxEnt method is 10--100 times faster than the BF method and grows linearly with respect to the number of edges (Fig.~\ref{fig:result_unweighted}A).
The networks sampled by all methods preserve the prescribed degree sequence comparably well (Fig.~\ref{fig:result_unweighted}B for three representative networks).
Quantitatively, the mean squared errors of the logarithmic degrees---defined as $N^{-1}\sum_{i=1}^N [\log(k_i+1) - \log(\hat k_i + 1)]^2$ for the degree $\hat k_i$ of a sampled network---are all comparable: 0.109$\pm$0.073, 0.109$\pm$0.073, and 0.109$\pm$0.072 for the BF, MH, and FastMaxEnt methods, respectively. Lastly, while the MH algorithm produces a rich-club structure substantially stronger than that of the BF, the profile for the FastMaxEnt method matches that of the unbiased network ensembles (Fig.~\ref{fig:result_unweighted}C).


\subsection{Weighted networks}
\label{sec:result_weighted}

\begin{figure}[tb!]
    \centering
    \includegraphics[width=\textwidth]{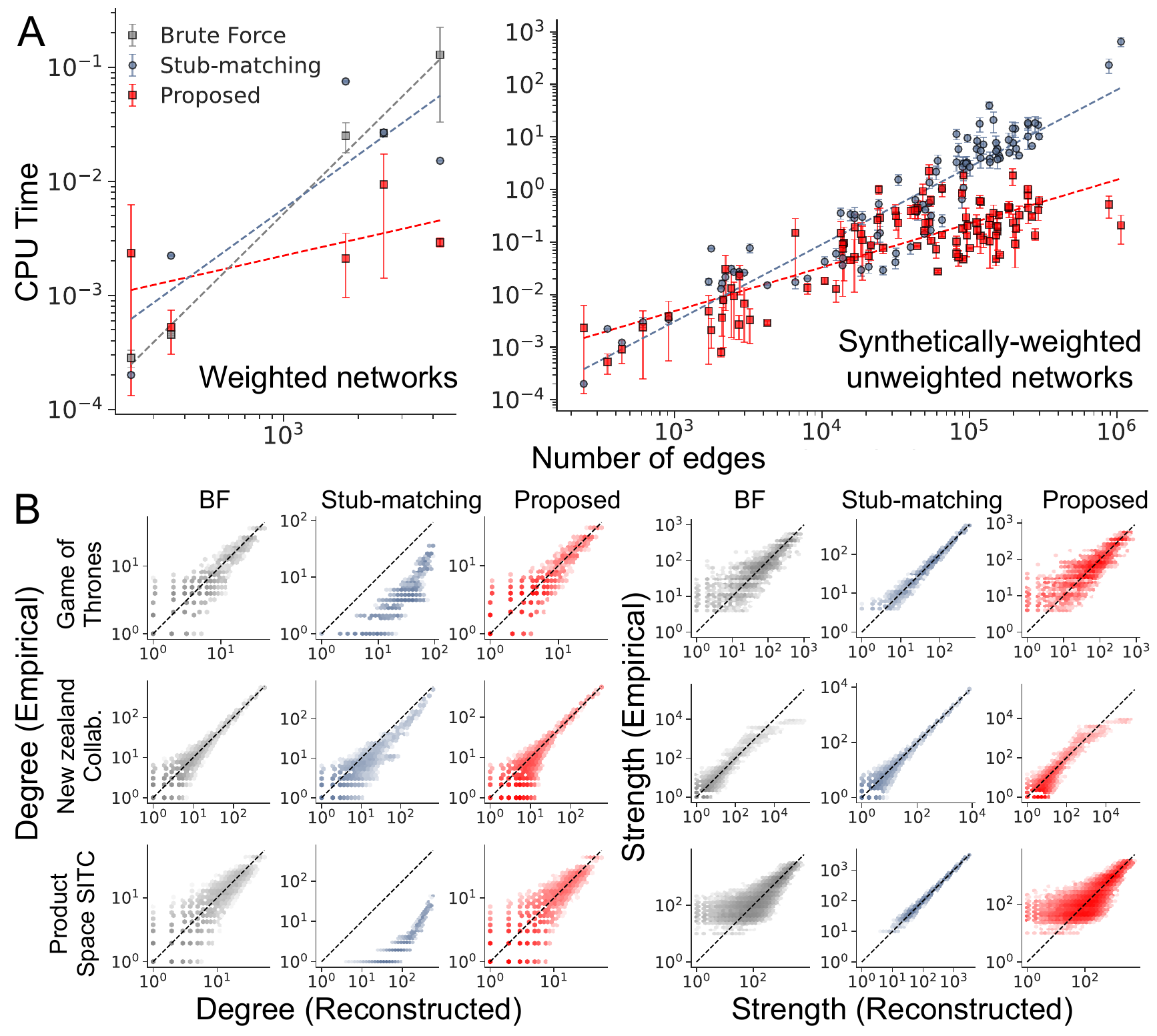}
    \caption{
    Results for the weighted networks.
    {\bf A}: CPU Time as a function of $M$.
    {\bf B}: The joint probability distribution of the reconstructed and empirical degree sequences, and that of the strength sequences.
    We show the results for three empirical networks as representative.
    }
    \label{fig:result_weighted}
\end{figure}

The FastMaxEnt method samples faster while being as unbiased as the BF method for both originally weighted and synthetically weighted networks (Fig.~\ref{fig:result_weighted}).
The FastMaxEnt method samples 10--100 times faster than the BF and strength-based sampling methods for large networks, with CPU time growing linearly with respect to the number of edges (Fig.~\ref{fig:result_weighted}A).
The BF and FastMaxEnt methods preserve the expected degree and strength relatively well, with a notable deviation of the strength sequence for the Product Space SITC network.
This is likely due to inference error in the parameters of the UECM, not the sampling algorithms per se, as the BF method also shows a similar deviation.
The strength-based sampling method based on the Chung-Lu model preserves the strength sequence well but falls short for the degree sequence.

\section{Discussion}
\label{sec:discussion}

We have developed a fast sampling algorithm for MaxEnt models by adapting the MH algorithm originally designed for the Chung-Lu model. Our evaluation demonstrated computational speedups of 10--100 times in CPU time for unweighted networks and 10--100 times for weighted networks relative to the brute-force method while preserving statistical accuracy.

Our approach echoes the combinatorial adaptation of the MH algorithm~\cite{sayama2018combinatorial} that replaces the underlying edge probability derived from combinatorial analysis of edge occurrences between two nodes. We did not compare against the combinatorial MH algorithm, as we focus on sampling efficiency rather than model differences. We note that both fast algorithms make it computationally feasible to systematically compare different configuration models on large, degree-heterogeneous networks where such comparisons were previously prohibitive.

Several limitations remain.
First, the inference cost for MaxEnt models is still ${\cal O}(N^2)$.
Second, we did not identify the exact theoretical computational complexity of our sampling algorithm, as it depends on the parameter distribution. The algorithm becomes most efficient when $\beta$ values are homogeneous, approaching $\beta_{\min} \simeq \beta_i + \beta_j$ for all node pairs $i, j$, as this ensures tight upper bounds throughout the sampling process.

Despite these limitations, our proposed sampling algorithm improves the efficiency of unbiased configuration models. We hope that our sampling algorithm bridges the gap between theoretical rigor and computational practicality.

\bibliographystyle{spmpsci} 
\bibliography{main} 

\end{document}